\documentclass[aps,preprint,showpacs,superscriptaddress]{revtex4}
\date{\today}
\usepackage{graphicx}

\begin{document}

\title{Shell-model study of spectroscopies and isospin structures in 
odd-odd $N=Z$ nuclei employing realistic \emph{NN} potentials}

\author{C. Qi}
\affiliation{School of Physics and MOE Laboratory of Heavy Ion
Physics, Peking University, Beijing 100871, China}
\author{F.R. Xu}
\email{frxu@pku.edu.cn} \affiliation{School of Physics and MOE
Laboratory of Heavy Ion Physics, Peking University, Beijing 100871,
China} \affiliation{Institute of Theoretical Physics, Chinese
Academy of Sciences, Beijing 100080, China} \affiliation{Center for
Theoretical Nuclear Physics, National Laboratory for Heavy Ion
Physics, Lanzhou 730000, China}

\begin{abstract}
The structures of odd-odd $N=Z$ nuclei in the lower \emph{fp} shell
have been investigated with a new isospin-nonconserving
microscopical interaction. The interaction is derived from a
high-precision charge-dependent Bonn nucleon-nucleon potential using
the folded-diagram renormalization method. Excellent agreements with
experimental data have been obtained up to band terminations.
Particularly, the relative positions of $T=0$ and $T=1$ bands were
well reproduced. Calculations with another interaction obtained from
the Idaho-A chiral potential showed similar results. Our
calculations also give a good description of the existence of
high-spin isomeric states in the $0f_{7/2}$ sub-shell. As examples,
the spectroscopies and isospin structures of $^{46}$V and $^{50}$Mn
have been discussed in detail, with the useful predictions of level
structures including electromagnetic properties. Results for
$^{50}$Mn were analyzed where experimental observations are still
insufficient.
\end{abstract}

\pacs{21.30.Fe, 21.60.Cs, 23.20.Lv, 27.40.+z}

\maketitle

\section{Introduction}
Nuclei with the equal numbers of neutrons and protons ($N=Z$) have
become a special ground to investigate neutron-proton (\emph{np})
correlations. In $N=Z$ nuclei, neutrons and protons occupy the same
shell-model orbits, leading to large spatial overlaps between
neutron and proton wave functions. The \emph{np} pairing is further
enhanced in odd-odd $N=Z$ nuclei due to odd neutron and odd proton
correlation. The \emph{np} residual pairing interaction can be
equally important as the neutron-neutron (\emph{nn}, $T_z=1, T=1$)
and proton-proton (\emph{pp}, $T_z=-1, T=1$) pairings. While
\emph{nn} and \emph{pp} pairings have been well established in terms
of the simple BCS model, the \emph{np} pairing is still an open
question that has motivated many recent theoretical and experimental
works \cite{Dean03,Schneider99,Poves98,Fries98,Schmidt00,Oleary02,
Brandolini01,Brentano01,Pietralla02,Svensson98,Sko98,
Lenzi99,OLeary99,Faes99,Satula01,Macchiavelli,Vogel00, Janecke05}.

In contrast to \emph{pp} and \emph{nn} correlations, the \emph{np}
pairing can form isovector ($T_z=0, T=1$) and isoscalar ($T_z=0,
T=0$) configurations. The isovector channel manifest itself in a
similar fashion to like-nucleon correlations, which can be evidenced
with the presence of nearly degenerate $T=1$ isobaric triplets along
the $N=Z$ line \cite{Macchiavelli}. The existence of the $T=0$
isoscalar component has also been shown theoretically, e.g., in Ref.
\cite{Satula01}. In light $N=Z$ nuclei, ground states usually favor
the lowest $T=0$ isospin quantum number, while $T=1$ states appear
at very high excitation energies. With increasing mass number,
however, $T=0$ and $T=1$ bands begin to coexist at low excitation
energies \cite{Dean03,Janecke05}. In particular, the isospin
structures of medium-mass \emph{fp} shell nuclei are very
interesting, where $T=1$ states can be lower than $T=0$ states.
$T=1$ ground states have been expected systematically
\cite{Janecke05,Vogel00}.

Recent advances in experimental techniques enable us to study the
structures of the odd-odd $N=Z$ nuclei in the \emph{fp} shell at
high spins and high excitation energies. This provides an unique
opportunity to investigate the interplay between $T=0$ and $T=1$
bands. Moreover, the high-precision measurements of level schemes
provide useful information for discussion of the isospin asymmetry
in their excited states at different angular momenta
\cite{Schneider99,Fries98,Schmidt00,Oleary02,Pietralla02,Svensson98,Sko98,
Lenzi99,OLeary99,Faes99}. Experimentally, isospin effects can be
discussed by defining the mirror energy differences (MED) and
triplet energy differences (TED) between analogous states in mirror
pairs and $T=1$ triplets, respectively
\cite{Lenzi01,Garrett01,Zuker02}. The differences have their origins
from the Coulomb interaction and charge dependence in strong force,
while the former usually plays a dominant role. In the \emph{fp}
shell, the coulomb shifts of single-particle orbits are relatively
small \cite{Zuker02}. This allows us to extract the effect from the
charge symmetry breaking in the strong force for the many-body
systems of nuclei.

The purpose of this paper is to study the structures and properties
of odd-odd $N=Z$ nuclei in the \emph{fp} model space by the
shell-model diagonalization method. The many-body interaction used
in the effective Hamiltonian has been derived from a high-precision
charge-dependent Bonn (CD-Bonn) nucleon-nucleon (\emph{NN})
potential \cite{Machleidt01} using the so-called folded-diagram
method \cite{Kuo90}. The obtained interaction retains the
charge-independence breaking effects in the \emph{NN} potential. The
contribution from the Coulomb field is also included. Hence the
effective Hamiltonian does not conserve isospin.

In Section II, we describe briefly the folded-diagram method and the
construction of the residual interaction. In Section III, the level
structures of the odd-odd $N=Z$ nuclei in the $fp$ shell are
calculated and discussed. In Section IV, we discuss the
electromagnetic properties of the odd-odd $N=Z$ nuclei with
analyzing the isospin structures of the nuclei.

\section{Effective interaction}

We start from the mean-field approximation with the perturbation
expansion. In this approximation, it is convenient to reduce the
degree of freedom of the Hilbert space for the many-fermion system
of nuclei. For large-scale shell-model diagonalizations, the
determination of the model space is a crucial step. The model space
can be defined with the projection operator,
\begin{equation}
P=\sum_{i=1}^D|\Phi_i\rangle\langle\Phi_i|,
\end{equation}
where $D$ is the dimension of the model space, and wave functions
$|\Phi_i\rangle$ are eigenfunctions of the unperturbed Hamiltonian,
$H_0=T+U$, with $T$ being the kinetic energy and $U$ an appropriate
one-body potential. The remaining space (so-called excluded space)
can be expressed by
\begin{equation}
Q=\sum_{i=D+1}^{\infty}|\Phi_i\rangle\langle\Phi_i|.
\end{equation}

The system of the full Hamiltonian with eigen energies $E_{\alpha}$
and wave functions $\Psi_{\alpha}$ that are determined by the full
Schr\"odinger equation of
$H|\Psi_{\alpha}\rangle=E_{\alpha}|\Psi_{\alpha}\rangle$ can be
reduced within the model space by projection \cite{Kuo90},
\begin{equation}\label{proj}
PH_{\text{eff}}P|\Psi_{\lambda}\rangle=(E_{\lambda}-E_c)P|\Psi_{\lambda}\rangle
=(E_{\lambda}-E_c)|\Phi_{\lambda}\rangle,
\end{equation}
with $E_{\lambda}\in \{E_{\alpha}\}~(\lambda=1,2,\cdots,D)$ and
$E_c$ is the energy of the assumed core left behind in practical
calculations. $H_{\text{eff}}$ is the mode-space-dependent effective
Hamiltonian. In the shell model context, the effective Hamiltonian
can be decomposed into two parts,
\begin{equation}
H_{\text{eff}}=H_{\text{0}}'+v_{\text{eff}},
\end{equation}
where
$H_{\text{0}}'=\sum_{\alpha}\varepsilon_{\alpha}'\text{a}^{\dag}_{\alpha}\text{a}_{\alpha}$
is the effective one-body Hamiltonian with $\varepsilon_{\alpha}'$
being single-particle energies (SPE). The $\varepsilon_{\alpha}'$
values can be obtained from the separation energies of valence
particles. Usually, the effective interaction $v_{\text{eff}}$ takes
the form of pure two-body interactions. It can be expressed as two
body matrix elements in harmonic oscillator (HO) basis.

The shell model provides the microscopic foundation to study the
excitations of nuclei. The Cohen-Kurath \cite{ck} and universal SD
(USD) \cite{USD} interactions have been proved to be quite
successful and useful for the $p$- and $sd$- shell calculations,
respectively. In past decades, extensive theoretical works have been
conducted to improve the calculations of the shell model for
$0f_{7/2}$ nuclei in which exact diagonalizations have become
possible due to the advance in computing techniques. The central
spirit of shell model calculations is to find a unified interaction
for a given mass region. For the \emph{fp} nuclei, various
interactions have been proposed, such as the Kuo-Brown (KB)
renormalized G matrix \cite{Kuo68} and its modified versions with
monopole centroid corrections \cite{KB3,Caurier05}, the FPD6
analytic two-body potential \cite{Richter91} and the GXPF
interaction \cite{Honma02}. These interactions have been extensively
tested with the successful descriptions of  rotational
collectivities of nuclei in the upper part of the $0f_{7/2}$ shell
(e.g., $^{48}$Cr) \cite{Caurier94,Caurier95,MPinedo96a}. However,
these interactions conserve isospin symmetry. In this work, we
present a new isospin-nonconserving effective interaction derived
microscopically from the underlying $NN$ potential.

Traditionally, effective residual interaction is derived from
phenomenal fit to observations \cite{USD} (e.g., the USD interaction
for the $sd$ shell.) or from the evaluation of the G reaction matrix
\cite{Caurier05}. Although the effective interaction can in
principle be derived from true $NN$ interaction, it has to be
renormalized to include the contributions from the excluded
configurations and to avoid the hard core of the bare $NN$
potential. Renormalization methods have been proposed for the
calculation of effective interaction from underlying meson-exchange
potentials, like the momentum-space renormalization group decimation
method ($V_{\text{low-}k}$ approach) \cite{Bogner03} and the unitary
correlation operator method (UCOM) \cite{Feldmeier98}. In
$V_{\text{low-}k}$ approach, high-momentum contributions are
integrated out with the introduction of a cutoff, while UCOM
describes short-range correlations by a unitary transformation. In
this paper, we used the so-called folded-diagram expansion method in
which it is more convenient to include effects from excluded
configurations. The folded-diagram expansion is a time-dependent
perturbation theory introduced by Kuo {\it et al.} for the
evaluation of residual interactions microscopically from bare
\emph{NN} potentials \cite{Kuo90,Jensen95}.  The short-range
repulsion is taken into account by the introduction of G matrix.

The folded (time-incorrect) and non-folded (time-correct) diagrams
are introduced in the evaluation of the time-evolution operator
U$(0,-\infty)$ which can be written using perturbation expansion in
the complex-time limit as,
\begin{widetext}
\begin{equation}
\text{U}(0,-\infty)=\lim_{t\rightarrow-\infty(\varepsilon)}\sum_{n=0}^{\infty}
\frac{1}{n!}(\frac{-\text{i}}{\hbar})^n\int_t^0\text{d}t_1 \int_t^0\text{d}t_2
\cdots\int_t^0\text{d}t_nT[H_1(t_1),H_1(t_2)\cdots H_1(t_n)],
\end{equation}
\end{widetext}
where
$\lim_{t\rightarrow-\infty(\varepsilon)}\equiv\lim_{\varepsilon\rightarrow
0^+} \lim_{t\rightarrow-\infty(1-\text{i}\varepsilon)}$ and \emph{T}
is the time-ordering operator in degenerate or nearly degenerate
systems. For a model-space with the dimension $D$, the lowest $D$
eigenstates of the true Hamiltonian can be constructed using the
time-evolution operator with model-space states, $\Phi_i$,
\begin{equation}
\lim_{t\rightarrow-\infty(\varepsilon)}\frac{
\text{U}(0,-\infty)|\rho_{\lambda}\rangle}{\langle\rho_{\lambda}|\text{U}(0,-\infty)|\rho_{\lambda}\rangle}
=\frac{|\Psi_{\lambda}\rangle}{\langle\rho_{\lambda}|\Psi_{\lambda}\rangle},
\end{equation}
where $\lambda=1,2,\cdots,D$ and $|\rho_{\lambda}\rangle$ is a set
of $D$ parent states defined as
\begin{equation}
|\rho_{\lambda}\rangle=\sum_{i=1}^DC_i^{(\lambda)}|\Phi_i\rangle,~~~~~\lambda=1,2,\cdots,D.
\end{equation}

In the folded-diagram decomposition procedure, the wave function
$\text{U}(0,-\infty)|\Phi_{i}\rangle$ can be factorized as
\begin{equation}
\text{U}(0,-\infty)|\Phi_{i}\rangle=
\text{U}_V(0,-\infty)|\Phi_{i}\rangle\times
\text{U}(0,-\infty)|c\rangle,
\end{equation}
where $|c\rangle$ denotes the core and the subscript V means that
all interactions in $ \text{U}_V(0,-\infty)$ are linked to valence
particles. Moreover, diagrams contained in
$\text{U}_V(0,-\infty)|\Phi_{i}\rangle$ can be expressed in terms of
so-called $\hat{Q}$-boxes. The $\hat{Q}$-box is made up of
non-folded diagrams which are irreducible and valence-linked
\cite{Kuo90}. It is defined as
\begin{equation}
\hat{Q}(\omega)=PH_1P+PH_1\mathcal{Q}\frac{1}{\omega-\mathcal{Q}H\mathcal{Q}}\mathcal{Q}H_1P,
\end{equation}
where $\omega$ is the energy of interacting nucleons. The exclusion
operator $\mathcal{Q}$ appears due to the Pauli principle preventing
interacting nucleons from scattering into states occupied already by
other nucleons \cite{Jensen95}.

Using the generalized time ordering, the effective interaction
$v_{\text{eff}}$ can be expanded in terms of $\hat{Q}$-boxes:
\begin{equation}\label{eff}
v_{\text{eff}}=\hat{Q}-\hat{Q}'\int\hat{Q}+\hat{Q}'\int\hat{Q}\int\hat{Q}-\cdots,
\end{equation}
where $\int$ denotes a folding sign imposing relative time
constraints. $\hat{Q}'$ is a set of $\hat{Q}$-boxes that have at
least two $H_1$ vertices attached to valence lines, satisfying
$\text{d}\hat{Q}'/\text{d}\omega=\text{d}\hat{Q}/\text{d}\omega$.
Correspondingly, the solution of Eq. (\ref{eff}) can be derived
using an iteration procedure:
\begin{equation}
v_{\text{eff}}^{(n)}=\hat{Q}+\sum_{m=1}^{\infty}\frac{1}{m!}
\frac{d^m\hat{Q}}{d\omega^m}\{v_{\text{eff}}^{(n-1)}\}^m,
\end{equation}
where the energy dependence of the $\hat{Q}$-box can be removed.

In practical evaluations, non-folded diagrams are calculated firstly
and summed to give the $\hat{Q}$-box. To treat the strong repulsive
core of the bare $NN$ potential which is unsuitable for perturbation
treatments, the Brueckner $G$ reaction matrix is usually used. The
$\hat{Q}$-boxes can be expressed in terms of G-matrix vertices. The
$G$ matrix is defined by the solution of the Bethe-Goldstone
equation \cite{Jensen95},
\begin{equation}
G=V+V\frac{\mathcal{Q}}{\omega-\mathcal{Q}H_0\mathcal{Q}}G,
\end{equation}
satisfying
\begin{equation}
\langle\Psi| G |\Psi\rangle=\langle\Psi|V|\Psi\rangle,
\end{equation}
where $\Psi$ is the correlated two-body wave function.

In the past decade, various high-precision phenomenological
meson-exchange potentials have been proposed with reasonable
descriptions of the charge symmetry breaking and charge independence
breaking \cite{Machleidt01}. Initiated by Weinberg \cite{Entem03},
the chiral perturbative \emph{NN} potential derived from the low
energy effective Lagrangians of the QCD model shows another
promising way. All the potentials have similar low-momentum
behavior. In the present work, the bare $NN$ potential $V$ is taken
to be a new charge-dependent version of the Bonn one-boson-exchange
potential as described in Ref. \cite{Machleidt01}. For further
studies, theoretical works have also been carried out with a N$^2$LO
chiral \emph{NN} potential by Idaho group \cite{Entem03}.

In the present work, we are interested in nuclei in the \emph{fp}
shell. The model space contains
$0f_{7/2},~1p_{3/2},~0f_{5/2},~1p_{1/2}$ sub-shells with the core of
$^{40}$Ca. The $\hat{Q}$-boxes are approximated to the third order.
Single-particle energies are taken from the experimental excitation
energies of the odd neutron in $^{41}$Ca \cite{Kuo68}. In the CD
Bonn potential, charge symmetry breaking ($V_{pp}\neq V_{nn}$) and
charge independence breaking ($V_{pp}\neq V_{pn}\neq V_{nn}$)
effects are embedded in all partial waves with angular momentum
$J\leq4$. The contribution from the Coulomb field is also added when
calculating the G matrix. The derived interaction is composed of
three parts: the proton-proton, neutron-neutron and proton-neutron
interactions. No specific mass dependence of the interaction is
considered.

\section{level structure}

We calculated the level structures of odd-odd $N=Z$ nuclei in the
lower part of \emph{fp} shell with the residual interaction derived
as above. The effective Hamiltonian is diagonalized with the
shell-model code OXBASH \cite{Oxbash,Brown01}. The Hamiltonian
matrix is given in proton/neutron formalism, with the neutrons and
protons being naturally treated as two kinds of particles.

\begin{figure*}
\includegraphics[scale=0.70]{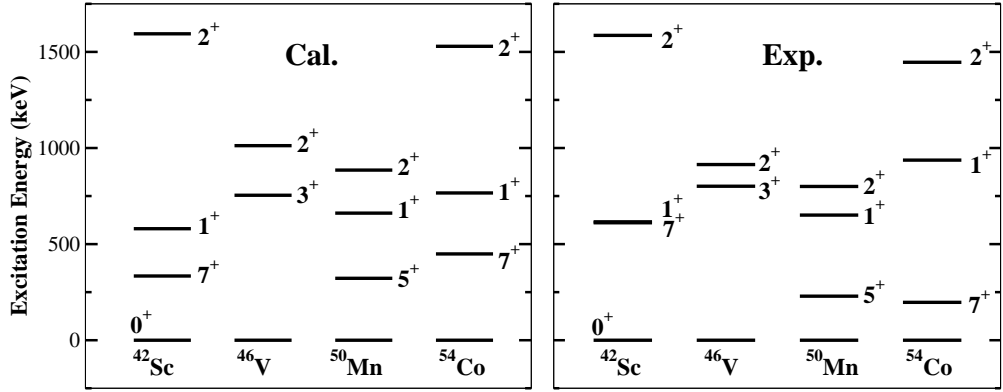}
\caption{\label{isomer}Low-lying states in odd-odd $N=Z$ nuclei in
the $0f_{7/2}$ shell. Calculated results and experimental data
\cite{toi98} are shown on the left-hand and right-hand side,
respectively. All $0^+$ ground states have the isospin of $T=1$. The
$2^+$ levels shown are the first excited $T=1$ states.}
\end{figure*}

The calculations of low-lying states in odd-odd $N=Z$ nuclei are
shown in Fig. \ref{isomer}. For nuclei with mass number $A\geq50$,
calculations have been done in truncated \emph{fp} model spaces.
While the $0^+$ and $2^+$ states have $T=1$, other states shown in
Fig. \ref{isomer} have $T=0$. Isospin is not an explicit quantum
number in our system. For a given state, the isospin quantum number
is assigned through the detailed comparison with neighboring
($T_z=\pm1$) nuclei. Odd-odd $N=Z$ nucleus and neighboring $|T_z|=1$
nuclei can form isospin triplet, leading to nearly identical $T=1$
bands. It can be seen that our calculations reproduce well the
relative positions of the $T=1$ and $T=0$ states. The largest
discrepancy compared with experiments is seen for the $7^+_1$ state
in $^{42}$Sc where the core breaking effect that are not taken into
account may be non-negligible. For the next heavier odd-odd $N=Z$
nucleus, $^{58}$Cu, our calculation can reproduce well the near
degenerating $T=1$ and $T=0$ bands, which are not shown in the
figure. For heavier odd-odd $N=Z$ nuclei, e.g. $^{62}$Ga,
contributions from the $0g_{9/2}$ sub-shell may become significant.

For comparison, we also used an interaction derived from the Idaho-A
chiral $NN$ potential \cite{Entem03} to calculate the low-lying
states of the nuclei in Fig. \ref{isomer}. Similar results are
obtained except $^{42}$Sc in which the calculated $7^{+}$ state
($T=0$) is lower than the $0^+$ state by about 47 keV in the Idaho-A
potential.

As mentioned before, for nuclei in $0f_{7/2}$ shell and above, $T=1$
ground states exist systematically. This situation has been studied
by Satu{\l}a and Wyss \cite{Satula01} with an extended mean-field
model in which the $T=1$ configuration is generated by isospin
cranking. Based on the analysis of symmetry energy coefficients, the
possible existence of isospin inversion has been extended to $A=96$
by J{\"a}necke and O'Donnell \cite{Janecke05}. In the shell-model
Hamiltonian, different components of two-body interactions are mixed
up. It is not easy to identify the role played by each component.
One possible way is to separate contributions from so-called
isovector and isoscalar pairing terms of the shell-model effective
interaction \cite{Poves98}. Such an effort has already been tried
\cite{Poves98}. In Ref.~\cite{Poves98}, authors argued that in the
$0f_{7/2}$ shell the competition between the isovector and isoscalar
terms can lower the centroid of the $T=1$ state, leading to a $T=1$
ground state. The shell model could provide an efficient and
qualitative way to search the interplay between various pairing
correlations on the isospin structures of self-conjugate nuclei. The
detailed evaluations of these terms would be done in future.

Tentative calculations have also been done for light nuclei in the
$p$- and $sd$- shell. Results for odd-odd $N=Z$ nuclei in these
shells are encouraging, showing that the folded-diagram method is
valid and efficient for the evaluation of shell-model interaction
from \emph{NN} potentials. It has long been a challenging problem
that previous realistic shell-model calculations cannot give the
right $3^+$ ground state in odd-odd $N=Z$ nuclei $^{10}_{~5}$B
\cite{Caurier05}. Our calculations in the \emph{psd} shell with the
CD-Bonn potential reproduce well the $3^+$ ground state and other
low-lying states.

\begin{figure}
\includegraphics[scale=0.40]{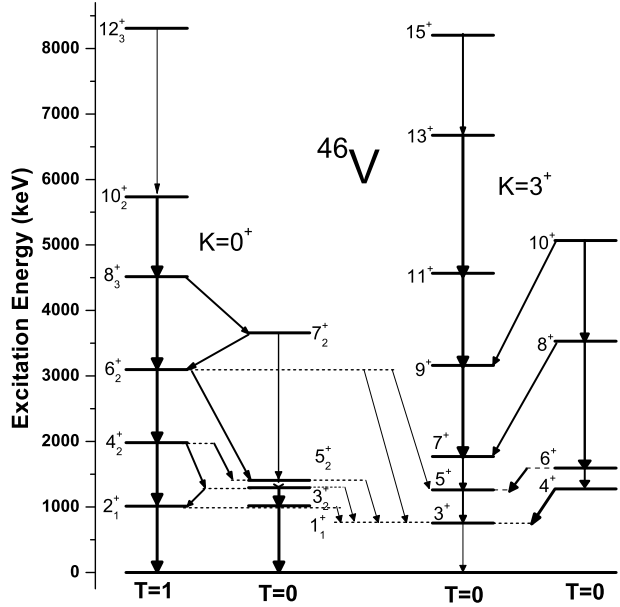}
\caption{\label{v46}Calculated level scheme in $^{46}$V. The
comparison with experimental observations is shown in Fig.
\ref{v46a}.}
\end{figure}

\begin{figure}
\includegraphics[scale=0.45]{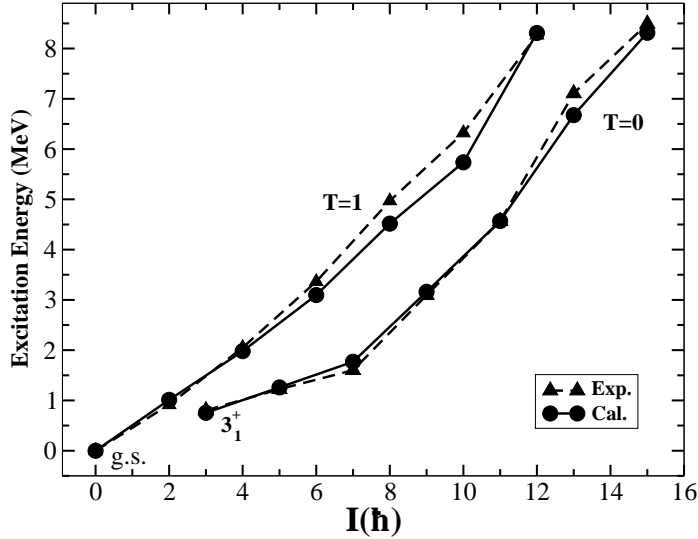}
\caption{\label{v46a}Comparison between calculated and experimental
$T=1$ and $T=0$ yrast bands in $^{46}$V. The experimental data are
taken from Refs. \cite{OLeary99,Garrett01}.}
\end{figure}

Experiments have observed high-spin states in $^{46}$V and $^{50}$Mn
up to the $0f_{7/2}$ band termination
\cite{Schmidt00,Oleary02,OLeary99,Svensson98}. Our calculations give
excellent agreement with experiments. Fig. \ref{v46} shows the
calculated level scheme for $^{46}$V. Comparisons with experiments
\cite{OLeary99,Garrett01} for $T=0$ and $T=1$ yrast bands are given
in Fig. \ref{v46a}. The ground-state band has been assigned with the
quantum number of $K^{\pi}=3^+$ and $T=0$ \cite{Oleary02}. Recent
experiments have deduced the $T=1$ ground-state bands in $^{46}$V
and its isospin-triplet partners, $^{46}$Ti and $^{46}$Cr, up to
$J^{\pi}=10^+$ \cite{Garrett01}. Our calculations can reproduce the
observations for the $T=1$ ground-state bands of these three nuclei.
In $^{46}$V, a state with the excitation energy of $E=8267$ keV was
detected and tentatively assigned as $12^+$ by Garrett {\it et al.}
\cite{Garrett01}. The state corresponds to our calculated $12^+_3$
state with the energy of 8309 keV in which the isospin quantum
number was assigned as $T=1$.

\begin{figure}
\includegraphics[scale=0.45]{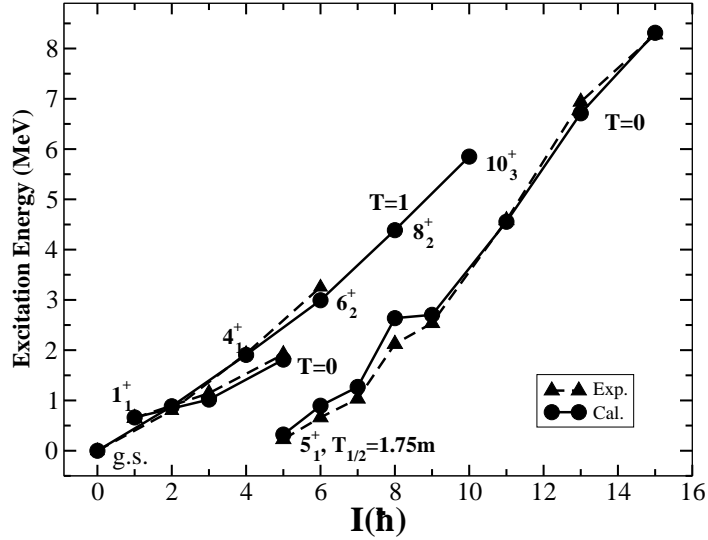}
\caption{\label{scheme}Calculated and experimental \cite{Oleary02}
lowest $T=1$ and $T=0$ states in $^{50}$Mn. The $T=0$ states from
$5^+$ to $15^+$ forms the $K^{\pi}=5^+$ band.}
\end{figure}

For the next heavier $N=Z$ nucleus, $^{50}$Mn, we chose a truncation
by allowing a maximum of two particles being excited to higher
sub-shells above the $f_{7/2}$ orbit. The truncation is valid since
configurations for low-lying states are governed by the $0f_{7/2}$
sub-shell. Fig. \ref{scheme} gives calculated and experimental
positive-parity states in $^{50}$Mn. The experiment \cite{Oleary02}
has identified the $T=0$ band up to the $0f_{7/2}$ band termination
and the $T=1$ band up to $6^+$. The observed $8^+$ state at 4874 keV
and $10^+$ state at 6460 keV were assigned $T=0$ by O'Leary {\it et
al.} \cite{Oleary02}. Our calculations reproduce the $T=0$ and $T=1$
states. $T=1$ states can be selected from the calculations of
electrical quadruple (E2) and magnetic dipole (M1) transition
properties that will be discussed in the next section.

\begin{figure}
\includegraphics[scale=0.50]{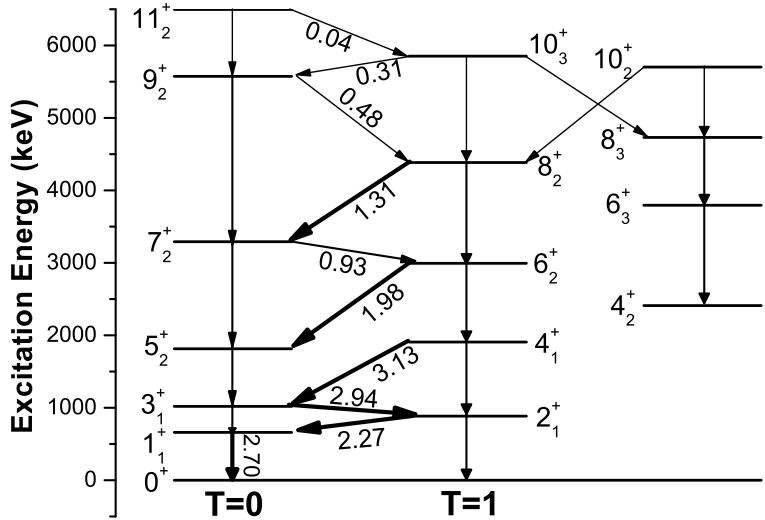}
\caption{Proposed isovector M1 transitions between the two $K=0^+$
bands in $^{50}$Mn. Calculated decay strengths (in $\mu_N^2$) are
shown along with the transition arrows. \label{emb}}
\end{figure}

In the framework of the deformed Nilsson diagram, low-energy
configurations in $^{50}$Mn have odd proton and odd neutron
occupying the $5/2^-[312]$ orbit of the $0f_{7/2}$ sub-shell,
leading to low-lying $K^{\pi}=0^+$ ($T=0$ or 1) and $K^{\pi}=5^+$
($T=0$) configurations that have been identified experimentally
\cite{Schmidt00,Pietralla02}. Our calculations have well reproduced
the observed states, as shown in Fig. \ref{scheme}. Experiments
observed that the $K^{\pi}=5^+$ state ($T=0$) is a $T_{1/2}=1.75$ m
isomer with an excitation energy of 229 keV
\cite{Schmidt00,Pietralla02}. Our calculation gives a energy of 322
keV for the isomer. Its decay is highly hindered due to the $K$
selection rule. Low-lying metastable $T=0$ states also exist in
other odd-odd $N=Z$ nuclei shown in Fig. \ref{isomer}, due to large
angular momentum gaps between $0^+$ ground states and low-lying
high-spin states. The $K^{\pi}=0^+$ band ($T=0$) in $^{50}$Mn has
been observed with three members of $1^+_1$, $3^+_1$ and $5^+_2$,
starting at the excitation energy of 651 keV for the $1^+$ state. In
present calculations, the lowest $1^+$ state appears at 661 keV,
forming a band as shown in Fig. \ref{emb}. Also from the properties
of inter-band M1 transition properties (discussed later), it can be
obtained that the calculated band built on the $1^+$ state
corresponds for the observed $K=0^+$ band. The recent experiment
\cite{Oleary02} assigned new quantum numbers, $J^{\pi}=5^+$ and
$T=0$, for the observed state at 1917 keV. Our calculation confirms
the $5^+_2$ state with a calculated energy of 1813 keV.

Three negative-parity states with possible $8^-,~10^-$ and $12^-$
assignments were observed in $^{50}$Mn \cite{Svensson98}. This band
has been tentatively interpreted as an octupole vibration band
coupled to the $5^+_1$ isomeric state by Svensson {\it et al.}
\cite{Svensson98}. In the shell model, the negative-parity band
could be generated in the $sdfp$ shell with one particle exciting
from the $0d_{3/2}$ orbital.

\section{Electromagnetic properties}

The electromagnetic (EM) transition properties of excited states can
provide other detailed structure information. Particularly, EM
properties play important roles in the identifications of various
bands. In the following, we will investigate E2 and M1 transition
properties in $^{46}$V and $^{50}$Mn, with a detailed analyse for
$^{50}$Mn.

In practical shell-model calculations, transitions are treated in
the truncated configuration. The wave functions ($\Phi_{\lambda}$)
of nuclear states are in fact the projections of the
full-Hamiltonian wave functions ($\Psi_{\lambda}$) onto the model
space with introducing an effective interaction as described by Eq.
(\ref{proj}). Hence, renormalized EM transition operators should be
used to include effects from the excluded space,
\begin{equation}
\langle \Psi_f|\mathcal{O}|\Psi_i\rangle\longrightarrow  \langle
\Phi_f|\mathcal{O}_{\text{eff}}|\Phi_i\rangle,
\end{equation}
where $i$ and $f$ indicate the initial and final states.

\begin{table*}
\centering \caption{Calculated and observed B(E2) values (in
e$^2$fm$^4$) in $^{50}$Mn. Two groups of effective charges are used.
One is the ``standard" charges with $e_p$=1.50e, $e_n$=0.50e; the
other one with $e_p$=1.15e, $e_n$=0.80e taken from Ref.
\cite{duRietz04}. Experimental data are taken from Ref.
\cite{Pietralla02}.}\label{be2}
\begin{ruledtabular}
\begin{tabular}{cccccccc}
&\multicolumn{3}{c}{$e_p$=1.50e, $e_n$=0.50e}&\multicolumn{3}{c}{$e_p$=1.15e, $e_n$=0.80e} &\\
$(J_i,T_i)\rightarrow (J_f,T_f)$ &HO &WS&SKcsb&HO & WS &SKcsb&EXP\\
\hline
\multicolumn{8}{l}{~~~~~~~$K=5^+$}\\
$(7_1^+,0) \rightarrow (5_1^+,0)$ & 45&42& 37&42&39&35&\\
$(9_1^+,0) \rightarrow (7_1^+,0)$ & 151 &143&126&142&133&118&$\geq 115^{+23}_{-19}$\\
$(11_1^+,0) \rightarrow (9_1^+,0)$ & 159&151&133&150&141&125& \\
$(13_1^+,0) \rightarrow (11_1^+,0)$ & 119&114&101&114&107&95& \\
$(15_1^+,0) \rightarrow (13_1^+,0)$ & 58 &55&49&55&52&46&\\
\multicolumn{8}{l}{~~~~~~~$K=0^+$}\\
$(3_1^+,0) \rightarrow (1_1^+,0)$ & 214& 205&178&203&191&167&$700^{+240}_{-190}$\\
$(5_2^+,0) \rightarrow (3_1^+,0)$ & 189&182& 158&180&170&148&$<1880$\\
$(7_2^+,0) \rightarrow (5_2^+,0)$ & 168&163&141&160&151&132& \\
$(9_2^+,0) \rightarrow (7_2^+,0)$ & 85&83&71&81&77&67& \\
$(11_2^+,0) \rightarrow (9_2^+,0)$ & 4.5&4.4 &3.9&4.3&4.1&3.7&\\
\end{tabular}
\end{ruledtabular}
\end{table*}


\begingroup
\squeezetable
\begin{table}
\caption{Calculated reduced transition strengths of M1  (in
$\mu^2_N$) and E2 (in e$^2$fm$^4$)  transitions in $^{46}$V.
Experimental data are taken from Refs.
\cite{Brandolini01,Brentano01}.}\label{v46bem}
\begin{ruledtabular}
\begin{tabular}{lccccccc}
&\multicolumn{2}{c}{B(E2)}&&\multicolumn{2}{c}{B(M1)}\\
\cline{2-3}\cline{5-6}
$J_i\rightarrow J_f $ ~~& ~~~~Cal.~~~~&~~~~Exp.~~~~&~~~~~~~&~~~Cal.~~~~&~~~~Exp.~~\\
\hline
$K=3^+$&&&&&\\
$5_1^+ \rightarrow 3_1^+$ & 29& 67(14)&&&\\
$7_1^+ \rightarrow 5_1^+$ & 2 &97(15)&&&\\
$9_1^+ \rightarrow 7_1^+$ & 173 &159(34)&&&\\
$11_1^+ \rightarrow 9_1^+$ & 153 &139(26)&&&\\
$13_1^+ \rightarrow 11_1^+$ & 109 &78(16)&&&\\
$15_1^+ \rightarrow 13_1^+$ & 78 &60(12)&&&\\
$4_1^+ \rightarrow 3_1^+$ &280 &208(50)&&$4\times10^{-4}$&\\
$6_1^+ \rightarrow 5_1^+$ & 157 &&&$9\times10^{-4}$&\\
$8_1^+ \rightarrow 7_1^+$ & 33 &&&$2\times10^{-7}$&\\
$10_1^+ \rightarrow 9_1^+$ & 33 &&&$3\times10^{-5}$&\\
$6_1^+ \rightarrow 4_1^+$ &96 &&&&\\
$8_1^+ \rightarrow 6_1^+$ & 138 &&&&\\
$10_1^+ \rightarrow 8_1^+$ & 130 &&&&\\
$K=0^+$&&&&&\\
$2_1^+ \rightarrow 0_1^+$ &176 &138(35)&&&\\
$4_2^+ \rightarrow 2_1^+$ & 237 &$>169$&&&\\
$6_2^+ \rightarrow 4_2^+$ & $221$ &&&&\\
$8_3^+ \rightarrow 6_2^+$ &212 &&&&\\
$10_2^+ \rightarrow 8_3^+$ & 161 &&&&\\
$12_3^+ \rightarrow 10_2^+$ & 0.03 &&&&\\
$3_2^+ \rightarrow 1_1^+$ &217 &&&&\\
$5_2^+ \rightarrow 3_2^+$ & 117 &&&&\\
$7_2^+ \rightarrow 5_2^+$ & 0.07 &&&&\\
$\Delta K=0,~~\Delta T=1$&&&&&\\
$1_1^+ \rightarrow 0_1^+$ & &&&1.49&$\geq0.77$\\
$3_2^+ \rightarrow 2_1^+$ &0.05&&&1.14&2.0(7)\\
$4_2^+ \rightarrow 3_2^+$ &0.15&&&1.0&0.57(15)\\
$4_2^+ \rightarrow 5_2^+$ & &&&1.26&0.55(13)\\
$6_2^+ \rightarrow 5_2^+$ &0.11 &&&0.33&\\
$7_2^+ \rightarrow 6_2^+$ & 0.12 &&& 0.03&\\
$8_2^+ \rightarrow 7_2^+$ & 3.7 &&& 0.06&\\
$\Delta K=3,~~\Delta T=1$&&&&&\\
$2_1^+ \rightarrow 3_1^+$ & 0.4 &&& 0.26&\\
$3_2^+ \rightarrow 3_1^+$ & 2.9 &&&2$\times10^{-5}$ &\\
$4_2^+ \rightarrow 3_1^+$ &  &&&0.04 &0.012\\
$4_2^+ \rightarrow 5_1^+$ &  &&&0.236 &0.02\\
$5_2^+ \rightarrow 3_1^+$ & 57 &&& &\\
\end{tabular}
\end{ruledtabular}
\end{table}
\endgroup

With the folded-diagram method, effective model-dependent transition
operators can de derived microscopically
\cite{Siiskonen01,Engeland00}. However, a simple way to evaluate the
renormalization effect is to introduce the effective charges of
nucleons \cite{Brown01}. For example, the effective operator for E2
transitions can be given as,
\begin{equation}
\mathcal{O}(\text{el}., 2M)=\sum_i(\tilde{e}_{\tau}/e)
r_i^2Y_2^M(\theta_i,\phi_i),\end{equation} where $i$ sums over all
particles and $\tilde{e}_{\tau}$ (with $\tau=\mu,~ \nu$) are
renormalized effective charges for protons and neutrons,
respectively.

Table \ref{be2} gives the shell-model predictions of B(E2) values
for $K^{\pi}=5^+$ and $K^{\pi}=0^+$ bands ($T=0$) in $^{50}$Mn. Two
different groups of empirical effective charges have been used. One
is the commonly-used effective charges of $e_p=1.5e$ and $e_n=0.5e$.
The other group with $e_p=1.15e$ and $e_n=0.80e$ was extracted from
the B(E2; $27/2^-\rightarrow 23/2^-$) values of $T=1/2$ mirror
nuclei $^{51}$Fe and $^{51}$Mn by du Rietz {\it et al.}
\cite{duRietz04}. The wave functions of the shell model are
calculated with three different bases, i.e., the harmonic oscillator
(HO), the Woods-Saxon (WS) potential and the Skyrme force (SKcsb) by
Brown {\it et al.} \cite{Brown00} in which a
charge-symmetry-breaking term has been added. In the $K=5^+$ band
($T=0$) of $^{50}$Mn, one E2 strength has been determined
\cite{Pietralla02}. The calculated B(E2, $9_1^+\rightarrow 7^+_1$)
value agrees with the measured lower limit. As seen from Table
\ref{be2}, calculated decay strengths are relatively stable for
different choices of effective charges and single-particle wave
functions. Reduced E2 decay strengths for $^{46}$V are shown in
Table \ref{v46bem} in which the results calculated with the HO
single-particle wave function and effective charges of e$_p$=1.5e
and e$_n$=0.5e are given.

\begin{figure}
\includegraphics[scale=0.51]{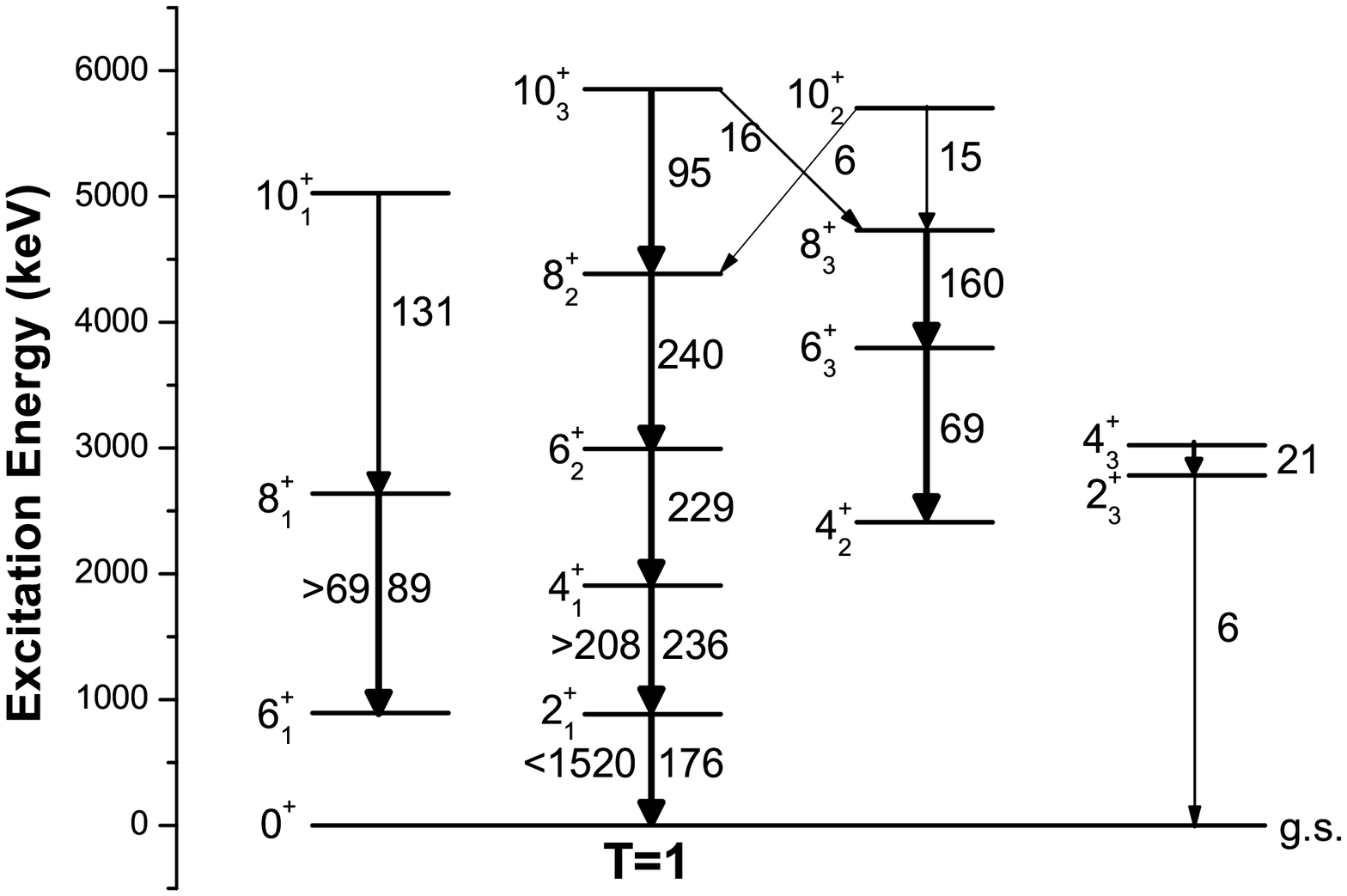}
\caption{\label{ema}Decay scheme and B(E2) strengths (in
e$^2$fm$^4$) of the proposed T=1 band in $^{50}$Mn. Calculated and
experimental strengths \cite{Pietralla02} are shown on the right-
and left-hand sides of the transition lines, respectively.  }
\end{figure}

Fig. \ref{ema} shows the calculations of electrical quadruple
transitions for the $T=1$ band in $^{50}$Mn. Results were calculated
with $e_p=1.5e$ and $e_n=0.5e$ and HO single-particle wave
functions. Three side bands identified experimentally are shown in
the figure. Two bands with $T=0$ on the right side of the figure
have been assigned with $K^{\pi}=3^+$ and $K^{\pi}=2^+$,
respectively, by Sahu {\it et al.} \cite{Sahu03}. The proposed
$10^+_1$ state ($T=0$) at 5027 keV, and the $8^+_2$ state ($T=1$) at
4386 keV have not been observed. Calculated $8^+_3$ and $10^+_2$
states are recognized as the observed $8^+$ state at 4874 keV and
the $10^+$ state at 6460 keV, respectively. Both states have the
isospin of $T=0$ since M1 transition strengths are rather weak
compared with inner-band E2 transitions.

The magnetic dipole (M1) operator is dominated by the isovector spin
($\sigma$) term. Strong M1 transitions have been observed only in
light nuclei. In the \emph{fp} shell, due to the coexistence of
$T=0$ and $T=1$ bands at low excitation energies, strong isovector
($\Delta T=1$) M1 transitions have been expected
\cite{Oleary02,Pietralla02}. We take the g-factor values of free
nucleon and the HO basis to investigate the isoscalar and isovector
M1 transition properties in $^{46}$V and $^{50}$Mn.

\begin{table}
\centering \caption{Calculated reduced matrix elements (in
$\mu^2_N$) for M1 transitions in $^{50}$Mn with the g-factors of
free nucleon [g$_{s}$(proton)=5.586, g$_s$(neutron)=-3.826,
g$_{l}$(proton)=1 and g$_{l}$(neutron)=0]. Experimental data are
taken from Ref. \cite{Pietralla02}. Calculated B(E2) values are also
shown.}\label{bm1}
\begin{ruledtabular}
\begin{tabular}{lccc}
&B(E2)&\multicolumn{2}{c}{B(M1)}\\
\cline{3-4}
$(J_i,T_i)\rightarrow (J_f,T_f) $&Cal.& Cal. &Exp.\\
\hline
\multicolumn{4}{l}{$\Delta T=1$}\\
$(1_1^+,0) \rightarrow (0_1^+,1)$ && 2.70& \\
$(2_1^+,1) \rightarrow (1_1^+,0)$ &0.28& 2.27 &$<$6.7\\
$(3_1^+,0) \rightarrow (2_1^+,1)$ &0.78$\times10^{-2}$& 2.94 &$2.9^{+10}_{-7}$\\
$(4_1^+,1) \rightarrow (3_1^+,0)$ &0.70& 3.13 &$1.05^{+31}_{-20}$\\
$(6_2^+,1) \rightarrow (5_2^+,0)$ &1.46& 1.98 &$>$0.24\\
$(7_2^+,0) \rightarrow (6_2^+,1)$ &0.30& 0.93 &\\
$(8_2^+,1) \rightarrow (7_2^+,0)$ &1.66& 1.31 &\\
$(9_2^+,0) \rightarrow (8_2^+,1)$ &0.93$\times 10^{-2}$& 0.48 &\\
$(10_3^+,1) \rightarrow (9_2^+,0)$ &0.43& 0.31 &\\
$(11_2^+,0) \rightarrow (10_3^+,1)$ &3.76& 0.04 &\\
\multicolumn{3}{l}{$\Delta T=0$}\\
$(4_2^+,0) \rightarrow (3_1^+,0)$ &2.0& 8.65$\times 10^{-5}$ &\\
$(4_3^+,0) \rightarrow (3_1^+,0)$ &0.34& 4.67$\times 10^{-2}$ &\\
$(8_3^+,0) \rightarrow (7_2^+,0)$ &7.5& $4.30\times 10^{-4}$ &\\
$(10_2^+,0) \rightarrow (9_2^+,0)$ &0.08& 0.023 &\\
$(11_2^+,0) \rightarrow (10_2^+,0)$&35 & 0.06 &\\
\end{tabular}
\end{ruledtabular}
\end{table}

Part of calculated M1 transition strengths for $^{46}$V and
$^{50}$Mn are shown in Table \ref{v46bem} and \ref{bm1},
respectively. Available experimental data are given for comparison.
Our calculations clearly show that $\Delta T=1$, M1 transitions with
$\Delta K=0$ are enhanced, while other $\Delta T=1$, M1 transitions
are retarded due to the \emph{K} quantum number selection (M1
transitions with $\Delta K \geq 1$ should be highly hindered). In
$^{46}$V, four strong isovector M1 transitions have been observed.
They were reproduced well in our calculations. In $^{50}$Mn, the
uncertainty in experimental data is very large \cite{Pietralla02}.
Calculated $\Delta K=0$ inter-band M1 transitions ($\Delta T=1$) are
shown in Fig. \ref{emb}. The strong M1 transitions have also been
interpreted by a simple picture of quasideuteron configurations of
odd neutron and odd proton in $^{50}$Mn \cite{Pietralla02}.

Another exciting problem involving odd-odd $N=Z$ nuclei is the study
of super-allowed decays. The $ft$ values of ground-state Fermi
($0^+\rightarrow 0^+$) decays would be the same for analogous
transitions if the conserved vector current (CVC) hypothesis is
valid \cite{Hardy05}. Odd-odd $N=Z$  nuclei provide the best ground
to test the understanding of the electroweak interaction. As
discussed in Ref. \cite{Hardy05}, the isospin-symmetry breaking
effect has significant contribution on the asymmetry in analogous
transitions. In our Hamiltonian, the isospin symmetry breaking is
treated microscopically, which can provide a basis to quantify the
effect. However, it is beyond the scope of this paper.

\section{Conclusion}

A microscopical \emph{fp}-shell interaction has been constructed
from the high-precision CD Bonn \emph{NN} potential using the
folded-diagram renormalization method. In the CD Bonn potential,
both charge symmetry breaking and charge independence breaking are
well described. The Coulomb field is also included in calculating
the effective interaction. Hence, in our effective Hamiltonian,
isospin symmetry is naturally broken. The isospin-nonconserving
interaction has been used to investigate the level structures of
odd-odd self-conjugate nuclei in the lower part of \emph{fp} shell.
Excellent agreements between calculations and experiments have been
obtained. Particularly, the interaction gave a good description of
the isospin structures in odd-odd $N=Z$ nuclei, reproducing well the
relative positions of $T=0$ and $T=1$ bands. Further studies done
with the Idaho-A chiral perturbation potential showed similar
results. Moreover, calculations show that effective interactions
derived with the folded-diagram method can reproduce the low-lying
structures of odd-odd $N=Z$ nuclei in other major shells, including
the famous case of $^{10}_{~5}$B in \emph{p} shell. Our calculations
reproduce the high-spin isomeric states in the \emph{fp} shell. In
particular, we have performed detailed shell-model calculations for
the positive-parity states in $^{46}$V and $^{50}$Mn. EM transition
properties and isospin structures in odd-odd $N=Z$ nuclei have been
predicted and discussed. We can construct a one-to-one
correspondence between calculated results and experimental data with
the comparisons of excitation energies and electromagnetic
transition properties. Our study shows that the predictions of this
\emph{ab initio} interaction is encouraging. It is very interesting
to test its performance in other \emph{fp} shell nuclei. Further
studies will be carried out in our future works.

\section*{Acknowledgement}
The authors are grateful to Prof. B.A. Brown for providing us the
code OXBASH on the request. This work has been supported by the
Natural Science Foundations of China under Grant Nos. 10525520 and
10475002, the Key Grant Project (Grant No. 305001) of Ministry of
Education of China. We also thank the PKU computer Center where
numerical calculations have been done.

\end{document}